\documentclass[11pt]{article}

\usepackage{textcomp}
\usepackage{textgreek}
\usepackage{natbib}
\usepackage{graphicx}
\usepackage{amsfonts,amssymb}
\usepackage{amsmath}
\usepackage{deluxetable}
\usepackage{upgreek}
\usepackage[auth-sc-lg]{authblk}
\usepackage[margin=1in,bottom=1.5in]{geometry}

\begin{document}

\title{Silicon isotopes reveal recycled altered oceanic crust in the mantle sources of Ocean Island Basalts}

\author{Emily A.\ Pringle}
\affil{Institut de Physique du Globe de Paris, Sorbonne Paris Cit\'{e}, Universit\'{e} Paris Diderot, CNRS, Washington University in St.\ Louis\\pringle@ipgp.fr}

\author{Fr\'{e}d\'{e}ric Moynier}
\affil{Institut de Physique du Globe de Paris, Sorbonne Paris Cit\'{e}, Universit\'{e} Paris Diderot, CNRS, Institut Universitaire de France}

\author{Paul S.\ Savage}
\affil{Institut de Physique du Globe de Paris, Durham University, University of St.\ Andrews}

\author{Matthew G.\ Jackson}
\affil{University of California, Santa Barbara}

\author{James M.D.\ Day}
\affil{Scripps Institution of Oceanography}

\date{\textit{Accepted to Geochimica et Cosmochimica Acta 4 June 2016}}

\maketitle

\begin{abstract}

The study of silicon (Si) isotopes in ocean island basalts (OIB) has the potential to discern between different models for the origins of geochemical heterogeneities in the mantle. Relatively large ($\sim$several per mil per atomic mass unit) Si isotope fractionation occurs in low-temperature environments during biochemical and geochemical precipitation of dissolved Si, where the precipitate is preferentially enriched in the lighter isotopes relative to the dissolved Si. In contrast, only a limited range ($\sim$tenths of a per mil) of Si isotope fractionation has been observed from high-temperature igneous processes. Therefore, Si isotopes may be useful as tracers for the presence of crustal material within OIB mantle source regions that experienced relatively low-temperature surface processes in a manner similar to other stable isotope systems, such as oxygen. Characterizing the isotopic composition of the mantle is also of central importance to the use of the Si isotope system as a basis for comparisons with other planetary bodies (e.g., Moon, Mars, asteroids).

Here we present the first comprehensive suite of high-precision Si isotope data obtained by MC-ICP-MS for a diverse suite of OIB. Samples originate from ocean islands in the Pacific, Atlantic, and Indian Ocean basins and include representative end-members for the EM-1, EM-2, and HIMU mantle components. On average, $\updelta$\textsuperscript{30}Si values for OIB ($-$0.32 $\pm$ 0.09\textperthousand{}, 2 sd) are in general agreement with previous estimates for the $\updelta$\textsuperscript{30}Si value of bulk silicate Earth \citep[$-$0.29 $\pm$ 0.07\textperthousand{}, 2 sd;][]{SAVAGE14}. Nonetheless, some small systematic variations are present; specifically, most HIMU-type (Mangaia; Cape Verde; La Palma, Canary Islands) and Iceland OIB are enriched in the lighter isotopes of Si ($\updelta$\textsuperscript{30}Si values lower than MORB), consistent with recycled altered oceanic crust and lithospheric mantle in their mantle sources. 

\end{abstract}

\section{Introduction}

Knowledge of Earth's internal structure and composition has primarily been obtained through complimentary approaches of geophysical and geochemical investigation. However, debate still persists about the mechanisms responsible for generating the chemical and physical characteristics of Earth's mantle observed today, as well as their evolution through time. The large-scale composition of the mantle may be inferred in part from mantle-derived melts erupted at or near the Earth’s surface, including mid-ocean ridge basalts (MORB) and ocean island basalts (OIB). Therefore, analyses of oceanic basalts remain a key component in understanding the geochemical complexities of the mantle.

Trace element ratios, radiogenic isotope variations (e.g., Sr, Nd, Hf, Os, Pb), and noble gas systematics have traditionally been studied in mantle-derived samples to better understand the chemical composition and physical processes occurring within the mantle throughout Earth history \citep[e.g.,][]{ZINDLER86,HOFMANN03,STRACKE05}. Ocean island basalts exhibit systematic differences in isotope compositions when compared with MORB; this has generally been attributed to mantle heterogeneities arising from the recycling of various materials into OIB mantle source regions \citep[e.g.,][]{HOFMANN82,HAURI93,HOFMANN97,GRAHAM02,JACKSON08,DAY09}. Various different mantle endmembers have been proposed to explain the variable Sr, Nd, and Pb isotope ratios, in that each plume samples chemically distinct mantle sources \citep{ZINDLER86}; this has been recently extended to Hf \citep[e.g.,][]{STRACKE12} and Os isotopes \citep[e.g.,][]{DAY13}. The main compositional mantle endmembers include DMM (Depleted MORB Mantle), EM-1 (Enriched Mantle 1), EM-2 (Enriched Mantle 2), and HIMU (“high-$\upmu$”, where $\upmu$ is defined as the time-integrated \textsuperscript{238}U/\textsuperscript{204}Pb). Each endmember is defined in terms of their relative radiogenic Sr and Pb isotope characteristics: DMM displays low \textsuperscript{87}Sr/\textsuperscript{86}Sr and relatively unradiogenic Pb, EM-1 has intermediate \textsuperscript{87}Sr/\textsuperscript{86}Sr and low \textsuperscript{206}Pb/\textsuperscript{204}Pb, EM-2 shows high \textsuperscript{87}Sr/\textsuperscript{86}Sr and intermediate \textsuperscript{206}Pb/\textsuperscript{204}Pb, and HIMU is defined by low \textsuperscript{87}Sr/\textsuperscript{86}Sr and high \textsuperscript{206}Pb/\textsuperscript{204}Pb \citep{ZINDLER86,HOFMANN03}. These geochemical characteristics are thought to be a reflection of specific lithologies present as a component of the source: EM-1 may incorporate recycled lower continental crust material or pelagic sediments \citep{WOODHEAD89,WILLBOLD10,GARAPIC15} or delaminated subcontinental lithosphere \citep{MCKENZIE83,MAHONEY91}, EM-2 may arise from the recycling of continental-derived sediments \citep{WHITE82,JACKSON07a}, and HIMU may involve recycled oceanic crust and lithospheric mantle \citep{HOFMANN82,CHAUVEL92,HAURI93,MOREIRAKURZ01,DAY10,CABRAL13}. It has even been proposed that an ancient marine carbonate component may exist in the HIMU mantle source \citep{CASTILLO15}. The traditional mantle endmembers represent extremes in terms of isotopic composition, with each OIB location potentially representing a unique mixture of mantle components. However, linking observations of chemical heterogeneity of OIB with the mechanisms responsible for generating such variations is currently a major challenge in mantle geochemistry.

In contrast to radiogenic isotope systems, the mantle end-members are less well defined by stable isotope variations; although O, Mg, and Ca show possible recycled signatures \citep[e.g.,][]{EILER01,HUANG11,WANG16}, generally it is assumed that the mantle is homogeneous with respect to most stable isotope systems. To date, this has been the case for silicon isotopes. For example, samples of and from the shallow mantle (peridotites and MORB) are assumed to be representative of the Si isotope composition of bulk silicate Earth (BSE), with any Si isotope heterogeneity being obliterated through convective mantle mixing \citep{SAVAGE10}. However, a systematic study of the Si isotope composition of OIB has not previously been performed; only a limited number of high-precision analyses are present in the literature \citep[see][for a recent review]{SAVAGE14}. Given the implications of possible mantle heterogeneities on a wide variety of mantle processes and the importance of an accurate characterization of the isotopic composition of BSE for geo- and cosmo-chemical comparison purposes, a dedicated investigation of potential Si isotope variations in OIB is warranted.

The Si isotope system is a potentially useful tool with which to investigate recycling of terrestrial surface materials within the mantle. Silicon is a key element in silicate minerals and is composed of three stable isotopes: \textsuperscript{28}Si (92.23\%), \textsuperscript{29}Si (4.68\%), and \textsuperscript{30}Si (3.09\%). Relatively large Si isotope fractionations have been observed as the result of low-temperature bio- and geo-chemical surficial/supergene processes, with the light Si isotopes becoming preferentially enriched in precipitated silica phases compared to dissolved Si in the fluid phase \citep{BASILE06}. Previous studies of the Si isotope systematics in low-temperature environments, including chert formation \citep{ROBERT06,BOORN07,BOORN10,TATZEL15}, diatom formation \citep{ROCHA97,HENDRY10}, and silicate weathering \citep{ZIEGLER05,OPFERGELT12}, have shown that Si isotope variations can range over several per mil in surface environments. In contrast, high-temperature igneous processes typically result in relatively small fractionations; terrestrial igneous materials display only a limited range ($\sim$0.2 per mil on the \textsuperscript{30}Si/\textsuperscript{28}Si ratio) of Si isotope compositions \citep{GEORG07,FITOUSSI09,CHAKRABARTI10,SAVAGE10,SAVAGE11,SAVAGE13a,SAVAGE13b,ARMYTAGE11, ZAMBARDI13,POITRASSON15}, with the bulk silicate Earth being relatively homogeneous \citep[$\updelta$\textsuperscript{30}Si = $-$0.29 $\pm$ 0.07\textperthousand{}, 2 sd;][]{SAVAGE14}. Partial mantle melting to form basaltic melts does not appear to generate large Si isotope fractionations \citep{SAVAGE10,SAVAGE11} compared to low-temperature processes, so incorporation of recycled surface materials in the mantle could produce Si isotope variations that may be observed in OIB.

Here we present high-precision Si isotope data obtained by multi-collector inductively-coupled-plasma mass-spectrometry (MC-ICP-MS) for a suite of OIB, including samples representative of the EM-1, EM-2, and HIMU mantle endmembers, to quantify potential Si isotope variations in the mantle and assess the processes that may generate Si isotope variability.

\section{Samples}

A set of 46 OIB representing locations in the Pacific, Atlantic, and Indian Oceans were selected for Si isotope analysis. In particular, samples were selected to represent the proposed mantle endmembers, including eight samples from Pitcairn (EM-1), eleven samples from Samoa (EM-2), and four samples from Mangaia (HIMU). Additional samples represent geographic locations with unique geochemical features, such as primitive noble gas compositions, or areas that display less extreme mantle endmember compositions. For example, three samples were selected from S\~ao Nicolau, one of the Northern Islands of Cape Verde that has been identified as HIMU-like based on radiogenic Pb isotope compositions and low \textsuperscript{87}Sr/\textsuperscript{86}Sr, but that also exhibits high \textsuperscript{3}He/\textsuperscript{4}He ($\sim10-12$ R/R$_{\textrm{A}}$), possibly indicative of a primitive lower mantle component \citep{DOUCELANCE03,MILLET08}. Similarly, two samples were selected from St. Helena, which has been identified as HIMU-like based on radiogenic isotope compositions but displays low \textsuperscript{3}He/\textsuperscript{4}He \citep{GRAHAM92,HANYU97}. Six samples were selected from the Canary Islands, particularly the islands of El Hierro and La Palma, whose mantle sources have been interpreted to contain distinct proportions of recycled oceanic crust and lithospheric mantle based on He--O--Sr--Nd--Os--Pb isotope systematics \citep{DAY09,DAY10,DAY11}. Four Icelandic samples from volcanic complexes in the Northern rift zone (NRZ) that display depleted tholeiitic compositions were selected. Iceland is one of the most well-studied OIB-type localities due to its association with both a spreading center (the Mid-Atlantic Ridge) and a deep-seated mantle plume \citep{RICKERS13}. The relatively unradiogenic He and Ne isotope ratios measured previously in these samples compared to global OIB is possible evidence for a deep, undegassed mantle source \citep{BREDDAM00,MOREIRA01}.

\begin{figure}
\centering
\includegraphics[totalheight=3in]{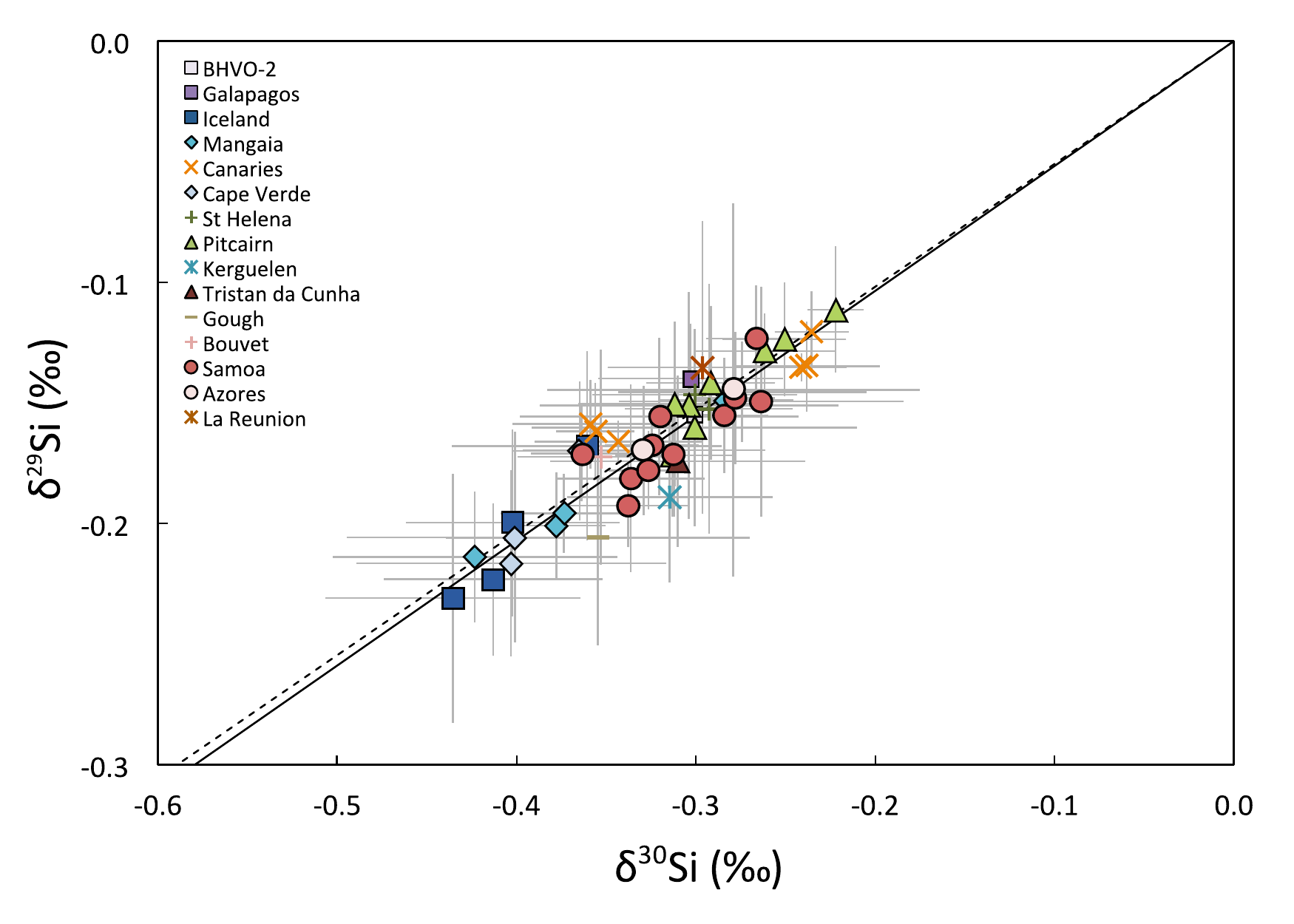}
\caption[Three-isotope plot of $\updelta$\textsuperscript{29}Si versus $\updelta$\textsuperscript{30}Si for OIB]{Three-isotope plot of $\updelta$\textsuperscript{29}Si versus $\updelta$\textsuperscript{30}Si for OIB. All data are within error ($\pm$ 2 se) of the calculated mass-dependent equilibrium (solid line, slope 0.5178) and kinetic (dashed line, slope 0.5092) fractionation lines.}
\end{figure}

\begin{deluxetable}{lllrrrrrrrr}
\tablewidth{0pt}
\tabletypesize{\tiny}
\tablecaption{Silicon isotope compositions of OIB and mineral separates relative to NBS28 as the bracketing standard.}
\tablehead{\colhead{Sample} & \colhead{Island} & \colhead{Ocean} & \colhead{$\updelta$\textsuperscript{29}Si} & \colhead{2 sd} & \colhead{2 se} & \colhead{$\updelta$\textsuperscript{30}Si} & \colhead{2 sd} & \colhead{2 se}& \colhead{n} & \colhead{SiO$_{2}$} }
\startdata
BHVO-2 & Hawaii & Pacific & $-$0.15 & 0.07 & 0.01 & $-$0.30 & 0.12 & 0.01 & 76 & 49.90 \\
AHANEMO2D20-B & Galapagos & Pacific & $-$0.14 & 0.06 & 0.02 & $-$0.30 & 0.13 & 0.05 & 6 & 48.89 \\
ISI09-9 & Iceland & Atlantic & $-$0.22 & 0.08 & 0.03 & $-$0.41 & 0.15 & 0.06 & 6 & 44.34 \\
KBD408702 & Iceland-Kambsfell & Atlantic & $-$0.23 & 0.14 & 0.05 & $-$0.44 & 0.19 & 0.07 & 7 & 47.76 \\
KBD408709 & Iceland-Kistufell & Atlantic & $-$0.20 & 0.09 & 0.04 & $-$0.40 & 0.13 & 0.06 & 5 & 48.26 \\
KBD408729 & Iceland-Blafjall & Atlantic & $-$0.17 & 0.08 & 0.04 & $-$0.36 & 0.15 & 0.08 & 4 & 51.19 \\
\textit{\textbf{Average-Iceland}} & & & \textbf{$-$0.21} & \textbf{0.06} & \textbf{0.03} & \textbf{$-$0.40} & \textbf{0.06} & \textbf{0.03} & \textbf{4} & - \\
MG-1001 \#1 & Mangaia & Pacific & $-$0.23 & 0.08 & 0.02 & $-$0.46 & 0.16 & 0.05 & 10 & 43.57 \\
MG-1001 \#2 & Mangaia & Pacific & $-$0.20 & 0.07 & 0.02 & $-$0.38 & 0.14 & 0.04 & 10 & 43.57 \\
\textit{Average-MG-1001} & & & $-$0.21 & 0.04 & 0.03 & $-$0.42 & 0.11 & 0.08 & 2 & 43.57 \\
MG-1002 \#1 & Mangaia & Pacific & $-$0.20 & 0.05 & 0.02 & $-$0.39 & 0.12 & 0.04 & 10 & 43.35 \\
MG-1002 \#2 & Mangaia & Pacific & $-$0.19 & 0.12 & 0.05 & $-$0.36 & 0.17 & 0.08 & 5 & 43.35 \\
\textit{Average-MG-1002} & & & $-$0.20 & 0.02 & 0.02 & $-$0.37 & 0.05 & 0.04 & 2 & 43.35 \\
MG-1006 \#1 & Mangaia & Pacific & $-$0.15 & 0.04 & 0.02 & $-$0.30 & 0.09 & 0.04 & 6 & 43.23 \\
MG-1006 \#2 & Mangaia & Pacific &  $-$0.14 & 0.04 & 0.02 & $-$0.26 & 0.07 & 0.03 & 6 & 43.23 \\
\textit{Average-MG-1006} & & & $-$0.15 & 0.01 & 0.01 & $-$0.28 & 0.05 & 0.04 & 2 & 43.23 \\
MG-1008 & Mangaia & Pacific & $-$0.20 & 0.05 & 0.02 & $-$0.38 & 0.07 & 0.03 & 6 & 43.00 \\
\textit{\textbf{Average-Mangaia}} & & & \textbf{$-$0.19} & \textbf{0.06} & \textbf{0.03} & \textbf{$-$0.36} & \textbf{0.12} & \textbf{0.06} & \textbf{4} & - \\
FE0903 & Canary-Fuerteventura & Atlantic & $-$0.12 & 0.04 & 0.02 & $-$0.24 & 0.05 & 0.02 & 6 & 47.05 \\
EH03 & Canary-El Hierro & Atlantic & $-$0.13 & 0.05 & 0.02 & $-$0.24 & 0.10 & 0.04 & 6 & 42.74 \\
EH12 & Canary-El Hierro & Atlantic & $-$0.14 & 0.01 & 0.01 & $-$0.24 & 0.06 & 0.03 & 6 & 42.08 \\
EH15 & Canary-El Hierro & Atlantic & $-$0.17 & 0.02 & 0.01 & $-$0.34 & 0.11 & 0.05 & 6 & 42.27 \\
\textit{\textbf{Average-El Hierro}} & & & \textbf{$-$0.15} & \textbf{0.04} & \textbf{0.02} & \textbf{$-$0.27} & \textbf{0.12} & \textbf{0.07} & \textbf{3} & - \\
LP03 & Canary-La Palma & Atlantic & $-$0.16 & 0.05 & 0.02 & $-$0.36 & 0.05 & 0.02 & 6 & 39.04 \\
LP15 & Canary-La Palma & Atlantic & $-$0.16 & 0.05 & 0.02 & $-$0.36 & 0.11 & 0.04 & 6 & 43.23 \\
\textit{\textbf{Average-La Palma}} & & & \textbf{$-$0.16} & \textbf{0.00} & \textbf{0.00} & \textbf{$-$0.36} & \textbf{0.00} & \textbf{0.00} & \textbf{2} & - \\
CV SN 98-01 & Cape Verde-Nicolau & Atlantic & $-$0.21 & 0.09 & 0.04 & $-$0.40 & 0.19 & 0.09 & 4 & 41.01 \\
CV SN 98-15 & Cape Verde-Nicolau & Atlantic & $-$0.17 & 0.06 & 0.03 & $-$0.36 & 0.08 & 0.04 & 4 & 39.50 \\
CV SN 98-18 & Cape Verde-Nicolau & Atlantic & $-$0.22 & 0.08 & 0.04 & $-$0.40 & 0.15 & 0.09 & 3 & 44.30 \\
\textit{\textbf{Average-Cape Verde}} & & & \textbf{$-$0.20} & \textbf{0.05} & \textbf{0.03} & \textbf{$-$0.39} & \textbf{0.04} & \textbf{0.02} & \textbf{3} & - \\
SH25 & St Helena & Atlantic & $-$0.15 & 0.12 & 0.05 & $-$0.29 & 0.10 & 0.05 & 5 & 44.95 \\
BM.1911,1626(1) & St Helena & Atlantic & $-$0.15 & 0.04 & 0.02 & $-$0.30 & 0.13 & 0.06 & 5 & 45.56 \\
\textit{\textbf{Average-St Helena}} & & & $-$0.15 & 0.01 & 0.01 & $-$0.30 & 0.01 & 0.01 & 2 &-  \\
PIT 1 & Pitcairn & Pacific & $-$0.13 & 0.04 & 0.02 & $-$0.26 & 0.10 & 0.04 & 6 & 48.10 \\
PIT 3 & Pitcairn & Pacific & $-$0.14 & 0.07 & 0.03 & $-$0.29 & 0.08 & 0.04 & 5 & 48.73 \\
PIT 4A & Pitcairn & Pacific & $-$0.12 & 0.06 & 0.02 & $-$0.25 & 0.08 & 0.03 & 6 & 48.91 \\
PIT 6 & Pitcairn & Pacific & $-$0.17 & 0.06 & 0.03 & $-$0.31 & 0.08 & 0.03 & 6 & 49.54 \\
PIT 7 & Pitcairn & Pacific & $-$0.15 & 0.12 & 0.05 & $-$0.30 & 0.20 & 0.08 & 6 & 48.15 \\
PIT 8 & Pitcairn & Pacific & $-$0.15 & 0.08 & 0.03 & $-$0.31 & 0.13 & 0.05 & 6 & 46.32 \\
PIT 12 & Pitcairn & Pacific & $-$0.16 & 0.10 & 0.04 & $-$0.30 & 0.22 & 0.09 & 6 & 49.22 \\
PIT 16 & Pitcairn & Pacific & $-$0.11 & 0.06 & 0.03 & $-$0.22 & 0.04 & 0.02 & 6 & 48.71 \\
\textit{\textbf{Average-Pitcairn}} & & & \textbf{$-$0.14} & \textbf{0.04} & \textbf{0.01} & \textbf{$-$0.28} & \textbf{0.07} & \textbf{0.02} & \textbf{8} & - \\
AG132 & Kerguelen & Indian & $-$0.19 & 0.06 & 0.04 & $-$0.31 & 0.10 & 0.06 & 3 & 45.08 \\
BM1962 128 (114) & Tristan da Cunha & Atlantic & $-$0.17 & 0.09 & 0.04 & $-$0.31 & 0.19 & 0.07 & 7 & 43.18 \\
ALR 40G & Gough & Atlantic & $-$0.21 & 0.09 & 0.04 & $-$0.35 & 0.17 & 0.08 & 4 & 48.54 \\
BV2 & Bouvet & Atlantic & $-$0.17 & 0.07 & 0.03 & $-$0.35 & 0.11 & 0.05 & 6 & 55.16 \\
OFU-04-14 & Samoa-Ofu & Pacific & $-$0.17 & 0.08 & 0.03 & $-$0.31 & 0.10 & 0.03 & 9 & 45.36 \\
ALIA-115-03 & Samoa-Savai'i & Pacific & $-$0.15 & 0.12 & 0.05 & $-$0.26 & 0.19 & 0.08 & 6 & 50.31 \\
ALIA-115-18 & Samoa-Savai'i & Pacific & $-$0.16 & 0.06 & 0.02 & $-$0.28 & 0.06 & 0.02 & 6 & 54.87 \\
T16 & Samoa-Ta'u & Pacific & $-$0.18 & 0.09 & 0.04 & $-$0.34 & 0.10 & 0.04 & 6 & 46.71 \\
T21 & Samoa-Ta'u & Pacific & $-$0.16 & 0.08 & 0.03 & $-$0.32 & 0.19 & 0.08 & 6 & 48.61 \\
T38 & Samoa-Ta'u & Pacific & $-$0.17 & 0.03 & 0.01 & $-$0.32 & 0.06 & 0.03 & 5 & 45.34 \\
74-4 & Samoa-Ta'u & Pacific & $-$0.18 & 0.04 & 0.02 & $-$0.33 & 0.05 & 0.02 & 5 & 45.52 \\
63-2 & Samoa-Vailulu'u & Pacific & $-$0.19 & 0.04 & 0.02 & $-$0.34 & 0.08 & 0.03 & 6 & 46.14 \\
76-9 & Samoa-Malumalu & Pacific & $-$0.17 & 0.05 & 0.02 & $-$0.36 & 0.07 & 0.03 & 6 & 45.91 \\
78-1 & Samoa-Malumalu & Pacific & $-$0.15 & 0.07 & 0.03 & $-$0.28 & 0.05 & 0.02 & 6 & 45.54 \\
128-21 & Samoa-Taumatau & Pacific & $-$0.12 & 0.05 & 0.02 & $-$0.27 & 0.07 & 0.03 & 6 & 49.82 \\
\textit{\textbf{Average-Samoa}} & & & \textbf{$-$0.16} & \textbf{0.04} & \textbf{0.01} & \textbf{$-$0.31} & \textbf{0.07} & \textbf{0.02} & \textbf{11} & - \\
ACO95-3 (1562 AD) & Azores-S\~ao Miguel & Atlantic & $-$0.17 & 0.06 & 0.03 & $-$0.33 & 0.15 & 0.07 & 5 & 47.81 \\
ACO2000-51 & Azores-S\~ao Jorge & Atlantic & $-$0.14 & 0.15 & 0.08 & $-$0.28 & 0.21 & 0.10 & 4 & 45.21 \\
\textit{\textbf{Average-Azores}} & & & \textbf{$-$0.16} & \textbf{0.04} & \textbf{0.02} & \textbf{$-$0.30} & \textbf{0.07} & \textbf{0.05} & \textbf{2} & - \\
Eruption 1931 & La Reunion & Indian & $-$0.14 & 0.14 & 0.06 & $-$0.30 & 0.12 & 0.05 & 5 & 43.91 \\
\textbf{Mineral separates} & & & & & & & & & & \\										
MG-1001 & zeolite & & $-$0.07 & 0.07 & 0.04 & 0.00 & 0.15 & 0.08 & 4 & 7.40 \\
MG-1001 & pyroxene & & $-$0.22 & 0.10 & 0.04 & $-$0.42 & 0.21 & 0.09 & 5 & 52.70 \\
\enddata
\end{deluxetable}

When possible, unaltered samples (based on visual inspection and absence of secondary minerals) were chosen for analysis in order to minimize the potential effects of post-eruptive weathering, which is a process that could alter the Si isotope composition of basalts. In many cases samples were fresh submarine or subglacial basaltic glasses. Major- and trace-element abundances and radiogenic isotopic ratios for the present sample set have been previously reported elsewhere \citep{LEROEX85,ALBAREDE88,BREDDAM00,CLAUDE01,BREDDAM02,WORKMAN04,GEIST06,JACKSON07a,JACKSON07b,MILLET08,KURZ09,DAY10,KAWABATA11,HART14,HERZBERG14,GARAPIC15}.  

\section{Analytical methods}

The alkali fusion sample dissolution and chemical purification methods used here have been described recently \citep{PRINGLE13a,PRINGLE13b,PRINGLE14,SAVAGE13}, and are based on the methods first described by \citet{GEORG06}. Samples were ground to a fine powder using an agate mortar; at least 1 g of each whole rock sample was processed to ensure representative sampling. Samples were then digested by an alkali fusion technique. This sample digestion method avoids the use of HF, which can lead to the loss of Si in the form of volatile SiF complexes following conventional acid digestion procedures. Approximately 10 mg of powdered sample was combined with $\sim${}200 mg NaOH flux in an Ag crucible and heated for 12 minutes in a muffle furnace at 720 C. The resulting fused sample was dissolved using Milli-Q H$_{2}$O and weakly acidified to 1\% HNO$_{3}$ to obtain a Si stock solution sufficiently dilute to avoid silicic acid polymerization. The samples were then purified for Si isotope analysis through ion-exchange chromatography using 1.8 mL BioRad AG50 X12 (200-400 mesh) cation exchange resin loaded in PolyPrep columns. Although the speciation of dissolved Si in water is pH dependent \citep{FUJII15}, it occurs as neutral or anionic species at low to neutral pHs and thus is not retained by the resin whereas cationic matrix elements are efficiently separated. Quantitative recovery of Si was achieved through elution by 5 mL Milli-Q H$_{2}$O. Column loads were calculated to yield final solution Si concentrations of 2 ppm for isotope analyses.

\begin{figure}
\centering
\includegraphics[totalheight=3.5in]{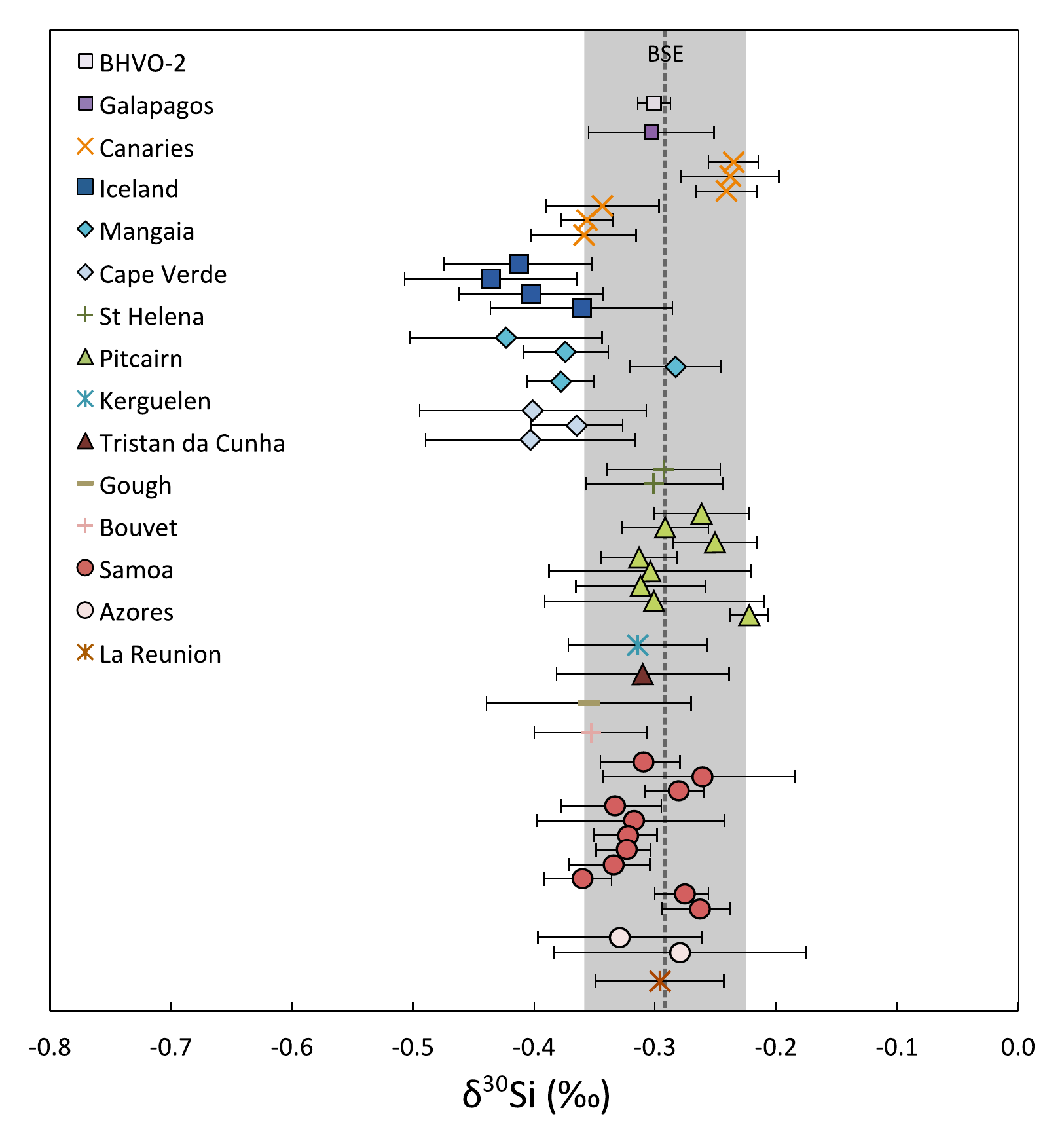}
\caption[Silicon isotope compositions of OIB relative to NBS28.]{Silicon isotope compositions of OIB relative to the bracketing standard NBS28 ($\pm$ 2 se). The dashed line and shaded box represent the estimate ($\pm$ 2 sd) for the Si isotope composition of the bulk silicate Earth (BSE) as given by \citet{SAVAGE14}. Most samples fall in the range of the estimate for BSE, but some HIMU and Iceland localities exhibit systematically lower $\updelta$\textsuperscript{30}Si values.} 
\end{figure}

Silicon isotope ratios were measured on Thermo Scientific Neptune Plus MC-ICP-MS instruments at Washington University in St. Louis and at the Institut de Physique du Globe de Paris. Each measurement represents 30 cycles with an 8.389 s integration time. Sample solutions were introduced using a wet plasma system and measurements were made in medium resolution on peak shoulder to avoid the large \textsuperscript{14}N\textsuperscript{16}O interference on mass 30. Standard--sample bracketing was used to correct for instrumental mass bias; the quartz sand standard NBS28 (NIST RM8546) and the well-characterized basalt geostandard BHVO-2 were each used as bracketing standards to obtain Si isotope measurements during separate analytical sessions and were subjected to the same dissolution and Si purification procedure as the samples.

Silicon isotope compositions are expressed relative to a bracketing standard (NBS28 or BHVO-2) using the following delta notation:
\begin{displaymath}
\updelta^{x}\text{Si}=\biggl(\frac{(^{x}\text{Si}/^{28}\text{Si})_{\text{sample}}}{(^{x}\text{Si}/^{28}\text{Si})_{\text{standard}}}-1\biggr)10^{3}
\end{displaymath}
where x = 29 or 30. Reported isotope ratios are averages of repeated measurements of each sample (``n'' in Table 1), and errors are determined from these repeated measurements. In a $\updelta$\textsuperscript{29}Si versus $\updelta$\textsuperscript{30}Si plot (Figure 1), all of the samples measured in this study fall within error of the calculated mass-dependent equilibrium (slope 0.5178) and kinetic (slope 0.5092) fractionation lines, and no Si isotope anomalies have been observed in terrestrial samples \citep{PRINGLE13b}. Therefore, the remainder of the text discusses the Si isotope data in terms of \textsuperscript{30}Si.

\section{Results}

Silicon isotope data relative to NBS28 as the bracketing standard are reported in Table 1 and Figure 2. The errors discussed throughout the text are 2 standard error (2 se) unless otherwise specified. To assess the reproducibility of the data, the Si isotope composition of the well-characterized basalt geostandard BHVO-2 was measured during each analytical session. The BHVO-2 value measured here ($\updelta$\textsuperscript{30}Si = $-$0.30 $\pm$ 0.01\textperthousand{}) represents 76 measurements during 13 separate analytical sessions and is in good agreement with previously published literature data \citep[e.g.,][]{ABRAHAM08,ARMYTAGE11,SAVAGE10,SAVAGE14}. 

On average, $\updelta$\textsuperscript{30}Si values for OIB ($-$0.32 $\pm$ 0.09\textperthousand{}, 2 sd), including the EM-1 endmember Pitcairn ($-$0.28 $\pm$ 0.07\textperthousand{}, 2 sd) and the EM-2 endmember Samoa ($-$0.31 $\pm$ 0.07\textperthousand{}, 2 sd), are in general agreement with previous estimates for the $\updelta$\textsuperscript{30}Si value of BSE \citep[$\updelta$\textsuperscript{30}Si = $-$0.29 $\pm$ 0.07\textperthousand{}, 2 sd;][]{SAVAGE14}. In addition, no variation in Si isotope composition with ocean basin is observable. However, some locations exhibit light Si isotope enrichments relative to MORB; on average, OIB from Iceland NRZ ($\updelta$\textsuperscript{30}Si = $-$0.40 $\pm$ 0.06\textperthousand{}, 2 sd) and the HIMU localities Mangaia (Cook Islands; $\updelta$\textsuperscript{30}Si = $-$0.36 $\pm$ 0.12\textperthousand{}, 2 sd), La Palma (Canary Islands; $\updelta$\textsuperscript{30}Si = $-$0.36 $\pm$ 0.00\textperthousand{}, 2 sd), and Cape Verde ($\updelta$\textsuperscript{30}Si = $-$0.39 $\pm$ 0.04\textperthousand{}, 2 sd) are lighter than the other OIB measured in this study (on average, $\updelta$\textsuperscript{30}Si = $-$0.30 $\pm$ 0.06\textperthousand{}, 2 sd). There are two notable exceptions within these localities with low $\updelta$\textsuperscript{30}Si. One sample from Mangaia (MG-1006) has anomalously high $\updelta$\textsuperscript{30}Si, but of the Mangaia group it has the lowest CaO/Al$_{2}$O$_{3}$ value. Additionally, one El Hierro sample (EH15) has high $\updelta$\textsuperscript{30}Si relative to the other El Hierro samples; this sample has the lowest $\updelta$\textsuperscript{18}O of the group.

\begin{deluxetable}{llrrrrrrr}
\renewcommand{\arraystretch}{1}
\tablewidth{0pt}
\tabletypesize{\scriptsize}
\tablecaption{Silicon isotope compositions of terrestrial samples relative to the basalt geostandard BHVO-2 as the bracketing standard.}
\tablehead{\colhead{Sample} & \colhead{Island} &\colhead{$\updelta$\textsuperscript{29}Si$_{\text{BHVO--2}}$} & \colhead{2 sd \tablenotemark{a}} & \colhead{2 se \tablenotemark{b}} & \colhead{$\updelta$\textsuperscript{30}Si}$_{\text{BHVO--2}}$ & \colhead{2 sd} & \colhead{2 se}& \colhead{n \tablenotemark{c}} }
\startdata
NBS28 & & 0.12 & 0.09 & 0.04 & 0.24 & 0.13 & 0.05 & 6 \\
ISI09-9 & Iceland & $-$0.05 & 0.05 & 0.02 & $-$0.10 & 0.10 & 0.04 & 6 \\
KBD408702 & Iceland & $-$0.03 & 0.07 & 0.03 & $-$0.09 & 0.12 & 0.05 & 6 \\
\textit{\textbf{Average-Iceland}} & & \textbf{$-$0.04} & \textbf{0.02} & - & \textbf{$-$0.10} & \textbf{0.01} & - & \textbf{2} \\
MG-1001 \#1 & Mangaia & $-$0.04 & 0.10 & 0.04 & $-$0.07 & 0.21 & 0.08 & 6 \\
MG-1001 \#2 & Mangaia & $-$0.05 & 0.14 & 0.05 & $-$0.10 & 0.20 & 0.07 & 6 \\
\textit{Average-MG-1001} & & $-$0.05 & 0.01 & 0.01 & $-$0.08 & 0.04 & 0.03 & 2 \\
MG-1002 \#1 & Mangaia & $-$0.02 & 0.06 & 0.02 & $-$0.01 & 0.08 & 0.03 & 6 \\
\textit{\textbf{Average-Mangaia}} & & \textbf{$-$0.03} & \textbf{0.03} & - & \textbf{$-$0.05} & \textbf{0.10} & - & \textbf{2} \\
SH25 & St Helena & 0.02 & 0.12 & 0.05 & $-$0.01 & 0.16 & 0.06 & 6 \\
PIT 1 & Pitcairn & 0.01 & 0.10 & 0.04 & 0.00 & 0.10 & 0.04 & 6 \\
OFU-04-14 & Samoa-Ofu & $-$0.02 & 0.07 & 0.03 & $-$0.03 & 0.12 & 0.05 & 6 \\
ACO2000-51 & Azores-S\~ao Jorge & $-$0.01 & 0.06 & 0.02 & $-$0.01 & 0.21 & 0.08 & 6 \\
\enddata
\tablenotetext{a}{2 sd = 2 $\times$ standard deviation}
\tablenotetext{b}{2 se = 2 $\times$ standard deviation/$\sqrt{}$n}
\tablenotetext{c}{n = number of measurements}
\end{deluxetable}

\begin{figure}
\centering
\includegraphics[totalheight=3.5in]{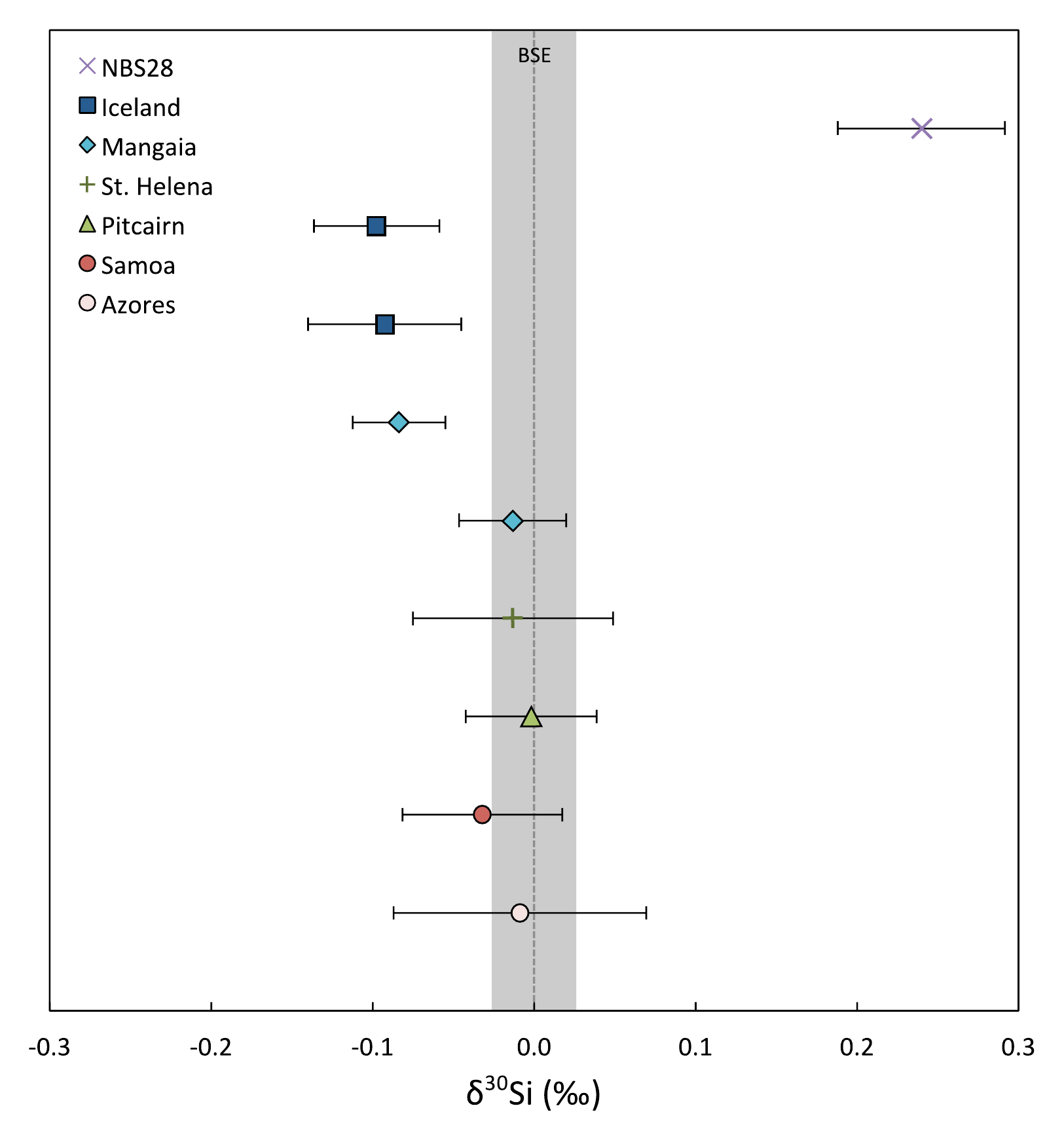}
\caption[Silicon isotope compositions of OIB relative to BHVO-2.]{Silicon isotope compositions of OIB relative to the basalt standard BHVO-2 as the bracketing standard ($\pm$ 2 se). The dashed line and shaded box represent the Si isotope composition of BHVO-2 ($\pm$ 2 se), which is used as a proxy for BSE. These additional data directly compare selected OIB samples to further distinguish Si isotope compositions and confirm the variations observed in Figure 2.} 
\end{figure}

In order to directly measure the variations in Si isotope composition among OIB, the basalt geostandard BHVO-2 was used as the bracketing standard in a separate measurement session; these data are denoted $\updelta$\textsuperscript{30}Si$_{\textrm{BHVO-2}}$ and are presented in Table 2 and Figure 3. BHVO-2 was chosen as a bracketing standard because its Si isotope composition is well characterized in the literature and it has previously been found to have the same Si isotope composition as the estimated value for BSE \citep{SAVAGE10,SAVAGE14}. BHVO-2 is an OIB from Kilauea volcano, taken from a surface layer of the pahoehoe lava that overflowed from the Halemaumau crater in Fall 1919 and has 7.23 wt.\% MgO. This direct comparison between OIB makes non-zero isotopic offsets easier to resolve since it eliminates a source of analytical error; when comparing data bracketed using the traditional NBS28 international Si isotope standard, the propagation of error compounds the uncertainty. When measured against BHVO-2, samples from Iceland ($\updelta$\textsuperscript{30}Si$_{\textrm{BHVO-2}}$ = $-$0.10 $\pm$ 0.01\textperthousand{}) and Mangaia ($\updelta$\textsuperscript{30}Si$_{\textrm{BHVO-2}}$ = $-$0.05 $\pm$ 0.10\textperthousand{}) are relatively lighter in Si isotopes, while samples from St. Helena, Pitcairn, Samoa, and the Azores are indistinguishable from BHVO-2 (on average, $\updelta$\textsuperscript{30}Si$_{\textrm{BHVO-2}}$ = $-$0.01 $\pm$ 0.02\textperthousand{}). These results are consistent with the data obtained through bracketing with the traditional Si isotope standard NBS28 and confirm the observed Si isotope variations in OIB.


\section{Discussion}

At first order, OIB exhibit generally uniform Si isotope compositions that are in good agreement with previous estimates for the $\updelta$\textsuperscript{30}Si value of BSE. However, small variations in the data are present, with some localities enriched in light Si isotopes by $\sim$0.1\textperthousand{} on $\updelta$\textsuperscript{30}Si compared to the estimate for BSE. Although this difference is small, it could have significant implications for mantle processes responsible for creating and preserving such heterogeneity. Below, we discuss the possible causes of the Si isotope variability observed in OIB, including post-eruptive alteration, magmatic differentiation, sampling of a primitive mantle reservoir, crustal assimilation, and recycling of crust and lithospheric mantle.

\subsection{Post-eruptive alteration}

Where possible, unaltered samples were chosen to minimize the effects of post-eruptive alteration. However, Mangaia volcanic rocks have crystallization ages of $\sim$20 Myrs and therefore have undergone surface weathering processes \citep{WOODHEAD96}. Although minor chemical weathering should not significantly affect the Si isotope composition of the bulk rock \citep{SAVAGE13a}, \citet{ZIEGLER05} found that basalt weathering may result in the formation of secondary minerals that are isotopically light relative to the dissolved Si in the fluid phase. To assess the possible contribution of isotopically fractionated secondary minerals on the bulk Si isotope composition of Mangaia samples, zeolite and pyroxene mineral separates from one Mangaia sample (MG-1001) were selected for Si isotope analysis by MC-ICP-MS. The Si isotope compositions of the MG-1001 mineral separates are reported in Table 1. The pyroxene fraction ($\updelta$\textsuperscript{30}Si = $-$0.42 $\pm$ 0.09\textperthousand{}) is indistinguishable from the MG-1001 whole-rock ($\updelta$\textsuperscript{30}Si = $-$0.42 $\pm$ 0.08\textperthousand{}) in Si isotope composition, suggesting that secondary minerals in this sample do not measurably affect the isotopic composition of the bulk rock.

In contrast, the zeolite separate ($\updelta$\textsuperscript{30}Si = 0.00 $\pm$ 0.08\textperthousand) is significantly heavier in Si isotope composition than the MG-1001 whole-rock, so the zeolites are not the carrier of the light Si isotope enrichment in the Mangaia OIB. However, this is unexpected considering that fluid alteration generally leads to secondary phases enriched in light Si isotopes. The heavier Si isotope composition of the zeolite can be explained as a precipitate that formed from Si-bearing surface fluids that were unassociated with dissolution of the host rock (i.e., fluids that do not have $\updelta$\textsuperscript{30}Si values matching those of basalts). Assuming a typical surface fluid Si isotope composition of 1.3\textperthousand{} \citep{ROCHA00,BOORN10} and considering that the Si isotope fractionation factor between precipitated and dissolved Si is around $-$1\textperthousand{} to $-$1.5\textperthousand{} for a variety of precipitates \citep[e.g.,][]{ROBERT06,STEINHOEFEL10}, the formation of secondary minerals with an isotopic composition around 0\textperthousand{} is possible if the zeolite Si source is dissolved Si in such a fluid. Isotopically-heavy silica coatings have been observed in weathered Hawaiian basalts and were interpreted as the result of weathering \citep{CHEMTOB15}. Since the zeolite fraction in MG-1001 has a relatively low modal abundance (\textless\ 10\%) and low concentration of Si (\textless\ 4 wt.\%), based on a simple mass balance the effect of the zeolite fraction on the bulk $\updelta$\textsuperscript{30}Si value is less than 0.04\textperthousand{}. This would result in a positive $\updelta$\textsuperscript{30}Si shift, suggesting that formation of secondary minerals due to weathering does not account for the light Si isotope enrichment in OIB lavas from Mangaia.

\subsection{Silicon isotope fractionation due to magmatic differentiation}

Since equilibrium isotope fractionation decreases with increasing temperature \citep[e.g.,][]{BIGELEISEN47}, only a limited range of Si isotope fractionation (on the sub permil level) is expected for high-temperature igneous processes. In general, evolved, silica-rich lithologies (e.g., rhyolites, dacites, granites) have heavier Si isotope compositions than less-evolved (ultramafic and basaltic) material, suggesting a small isotope fractionation during magmatic differentiation, with a $\sim$0.2\textperthousand{} offset in \textsuperscript{30}Si/\textsuperscript{28}Si from 40 to 75 wt.\% SiO$_{2}$ \citep{SAVAGE11,POITRASSON15}. First-principles calculations predict equilibrium Si isotope fractionation between co-existing mineral phases, with the more polymerized (i.e., more Si-rich) phase generally enriched in the heavier isotopes \citep{MEHEUT07}; this suggests that Si isotope fractionation may result from fractional crystallization or partial melting. However, no significant Si isotope fractionation has been observed as the result of basalt extraction by mantle partial melting \citep{SAVAGE10}.

\begin{figure}
\centering
\includegraphics[totalheight=3in]{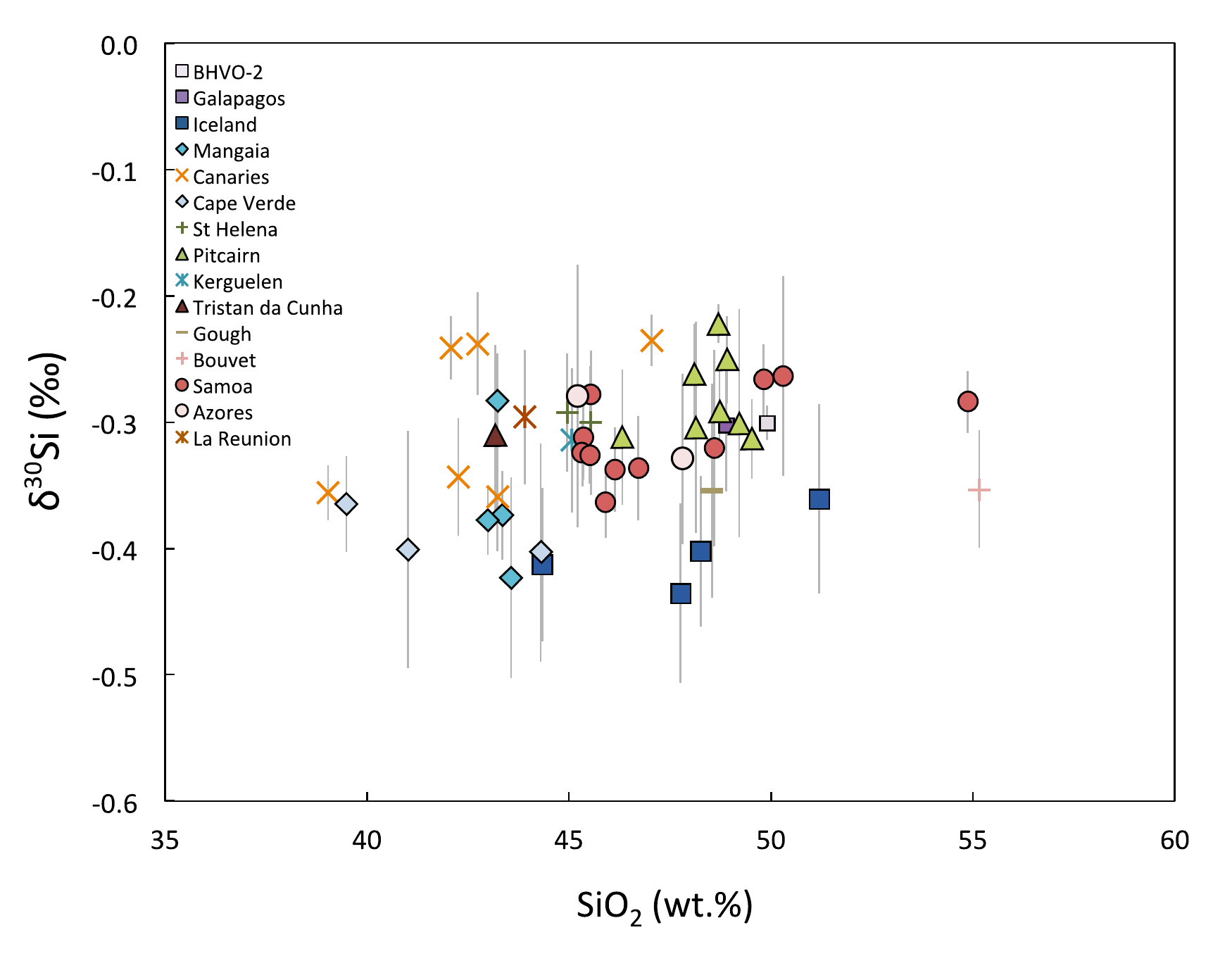}
\caption[Plot of $\updelta$\textsuperscript{30}Si as a function of SiO$_{2}$ for OIB.]{Plot of $\updelta$\textsuperscript{30}Si as a function of SiO$_{2}$ content for the OIB measured in this study. No correlation between Si isotope composition and silica content is observed, and widely variable $\updelta$\textsuperscript{30}Si values are observed for a given SiO$_{2}$ value. A plot of $\updelta$\textsuperscript{30}Si versus MgO or Mg\# shows similar results.} 
\end{figure}

The Si isotope compositions of OIB span a similar range (i.e., from the lightest OIB at $\updelta$\textsuperscript{30}Si = $-$0.44\textperthousand{} to the heaviest OIB at $\updelta$\textsuperscript{30}Si = $-$0.22\textperthousand{}) as that defined by ultramafic to rhyolitic material in the literature \citep[e.g.,][]{SAVAGE14}, but since the sample set presented here consists of basaltic rocks, there is a much smaller range of SiO$_{2}$ contents (Table 1). Furthermore, samples with similar SiO$_{2}$ contents (e.g., $\sim$48 wt.\% SiO$_{2}$) can span a range of Si isotope compositions up to 0.2\textperthousand{} (Figure 4). A plot of $\updelta$\textsuperscript{30}Si versus MgO or Mg\# shows a similar lack of correlation with $\updelta$\textsuperscript{30}Si; this further argues against accumulation of isotopically light olivine and/or pyroxene as the carrier of low $\updelta$\textsuperscript{30}Si values in OIB. The lack of correlation between $\updelta$\textsuperscript{30}Si and SiO$_{2}$ content suggests that magmatic differentiation is not the primary cause of Si isotope variability observed in OIB. 

\subsection{Preservation of primitive Si isotope heterogeneity in OIB source regions}

Recent theoretical calculations of equilibrium fractionation factors show that Si isotopes may be fractionated between silicate minerals at the elevated pressure conditions in the lower mantle \citep{HUANG14}. Due to the large pressure gradients within the deep mantle, high-pressure silicates have a different structure than those at lower pressure; low-pressure olivine polymorphs have four-fold coordination of Si while high-pressure bridgmanite (Mg-perovskite) has six-fold coordination of Si. This difference in Si coordination changes the vibrational frequency associated with the Si-O bonds, and isotopic substitution is sensitive to this effect. Although isotopic fractionations are generally small at high-temperature, the work of \citet{HUANG14} predicts significant Si isotope variations among silicate minerals with four-fold coordination of Si (e.g., olivine, pyroxene, garnet) and six-fold coordination of Si (bridgmanite). As a result, based on the calculated fractionation factors and estimates of the size and composition of mantle reservoirs, \citet{HUANG14} estimated the bridgmanite dominated lower mantle to be $\sim$0.1\textperthousand{} lighter on the \textsuperscript{30}Si/\textsuperscript{28}Si ratio compared to the upper mantle. 

If Si isotope heterogeneity was created by bridgmanite crystallization at the base of the magma ocean during Earth's silicate differentiation and subsequently preserved through mantle evolution, then the lower mantle may still be enriched in lighter Si isotopes compared to the upper mantle. Although there is much debate about the internal structure of the mantle and the mechanisms responsible for the long-term preservation of primitive mantle reservoirs, layered convection models suggest that geochemical heterogeneities may be preserved for long periods \citep[e.g.,][]{BRANDENBURG08,STIXRUDE12}.\

The degree of variation in Si isotope composition between OIB measured in this study and previously measured MORB derived from the shallow mantle \citep[e.g.,][]{SAVAGE10} is in agreement with the $\sim$0.1\textperthousand{} difference in Si isotope composition between the upper and lower mantle predicted by \citet{HUANG14}, suggesting that preservation of Si isotope heterogeneity within the mantle is a possible explanation for the $\updelta$\textsuperscript{30}Si variability in OIB. The variable Si isotope compositions of OIB could potentially reflect selective sampling of preserved primordial components with light Si isotope enrichment within the lower mantle. In the case of Icelandic NRZ OIB, their light Si isotope enrichments may be consistent with sampling of a deep, primitive, undegassed mantle source based on the He and Ne isotope systematics in these particular samples \citep{MOREIRA01}. Similarly, the S\~ao Nicolau island of Cape Verde displays both relatively primitive \textsuperscript{3}He/\textsuperscript{4}He \citep[10-15 R/R$_{\textrm{A}}$;][]{DOUCELANCE03} and a light Si isotope enrichment compared to MORB. 

Arguments against light Si isotope compositions reflecting a lower mantle signature include the observation that Samoan Ofu samples have primitive He \citep{JACKSON07b} and Ne \citep{JACKSON09} but MORB-like Si isotope composition ($\updelta$\textsuperscript{30}Si = $-$0.31 $\pm$ 0.03). Conversely, Canary Island samples have \textsuperscript{3}He/\textsuperscript{4}He values within the range of MORB \citep{DAY11}, but have variable Si isotope compositions. Therefore, sampling of a primitive mantle reservoir with a light Si isotope composition cannot solely explain the Si isotope variability in OIB but remains a possibility. Experimental work to constrain Si isotope fractionation between silicates with different Si coordination are required to ground-truth the theoretical calculations of \citet{HUANG14}. 

\subsection{Assimilation of crustal material}

Although only a limited number of samples in this study have previously been analyzed for O isotope compositions, it is noteworthy that most of the locations that exhibit light Si isotope enrichments also have been identified as areas with low $\updelta$\textsuperscript{18}O signatures, including HIMU-type OIB and Iceland \citep{DAY09,EILER01,HARMON95}. This raises the question of whether there is a common mechanism that generates the isotopically-light Si and O isotope compositions of OIB. Like Si isotopes, O isotopes display large variations in surface environments relative to mantle materials; therefore, O isotopes can be a sensitive measure of low-temperature processes, particularly water-rock interactions. Altered oceanic crust can be enriched in the heavier isotopes of O through low-temperature alteration by surface fluids (characteristic of the upper oceanic crust) or depleted in the heavier isotopes through high-temperature hydrothermal alteration (characteristic of the lower oceanic crust), so variations in $\updelta$\textsuperscript{18}O may indicate the presence of oceanic crustal material in OIB magmas. 

The depletions in the heavy isotopes of O found in some HIMU locations and Iceland are consistent with either recycling or assimilation of hydrothermally altered lower oceanic crust \citep{EILER01}. It is difficult at this point to distinguish between recycling and assimilation of oceanic crust based solely on Si isotopes. However, due to the much lower concentration of dissolved Si in fluids compared to the high O concentration, and the similar Si isotope composition of Si saturated hydrothermal fluids to their rock setting \citep{ANDRE06,ROCHA00}, the effect of hydrothermal alteration is likely to be much smaller for Si isotopes than for O isotopes. Large amounts of hydrothermally altered material would likely be needed to create resolvable Si isotope variations in OIB. 

Based on the comparison of $\updelta$\textsuperscript{18}O and \textsuperscript{187}Os/\textsuperscript{188}Os in OIB, \citet{EILER01} suggested that HIMU-type lavas compositions are inconsistent with the incorporation of recycled oceanic crust, and $\updelta$\textsuperscript{18}O depletions in OIB reflect assimilation of lower oceanic crust. However, \citet{DAY09} argued that radiogenic \textsuperscript{187}Os/\textsuperscript{188}Os coupled with high Os concentrations in some Canary Island OIB cannot be explained by crustal assimilation, but rather indicates a contribution of recycled oceanic lithosphere. This suggests that crustal assimilation is not the main cause of the light Si enrichments in this OIB suite, but processes of crustal recycling may play an important role in generating Si isotope variability in the mantle.

\subsection{Contribution of recycled material to the Si isotope variability observed in OIB}

Subduction introduces a wide variety of materials into the mantle, including oceanic and continental crust, oceanic lithospheric mantle, and sediments, all of which may have undergone variable degrees of alteration on the seafloor and/or during subduction \citep[e.g.,][]{HOFMANN97,HOFMANN82,PLANK98}. Subduction of geochemically diverse material that is incorporated into hotspot sources may give rise to observed heterogeneities in the products of these magmas, and it has been proposed that the type of material subducted generates the isotopic signatures of the different mantle components sampled by some OIB \citep{ZINDLER86}. Furthermore, subducted sediment is a significant source of incompatible elements to the mantle, and sediments may aid in creating chemical and isotopic heterogeneity within the mantle \citep{WHITE82,CHAUVEL07,JACKSON07a}. 

The presence in the mantle of unaltered continent-derived material (in the form of lower continental crust or sediments derived from the upper continental crust that has not been altered in Si isotope composition due to metamorphic dehydration and fluid-rock interaction during subduction) is unlikely to significantly affect OIB Si isotope compositions. Estimates of the Si isotope composition of both the lower and upper continental crust ($\updelta$\textsuperscript{30}Si = $-$0.29\textperthousand{} and $\updelta$\textsuperscript{30}Si = $-$0.25\textperthousand{}, respectively) are similar to MORB and estimates for the BSE \citep{SAVAGE13a,SAVAGE13b}. Although specific continental inputs may have Si isotope compositions that differ from these values for the lower and upper continental reservoirs \citep[e.g., shale, which exhibits $\updelta$\textsuperscript{30}Si values from zero to $-$0.82\textperthousand{};][]{SAVAGE13a}, we assume that on a large scale a single rock type would not dominate the subducted material and such sub-permil $\updelta$\textsuperscript{30}Si variations would be averaged out. Similarly, the presence of unaltered oceanic crust (with MORB-like Si isotope composition) would not create Si isotope variations in OIB. However, isotopic variations in HIMU-like OIB and the Iceland NRZ OIB in this study have most often previously been explained by the incorporation of recycled oceanic crust into the plume source \citep[e.g.,][]{HOFMANN82,HAURI93,WOODHEAD96,CHAUVEL00,MOREIRA01,BREDDAM02,DOUCELANCE03,MACPHERSON05,THIRWALL06,MILLET08,DAY09,KAWABATA11,CABRAL13,HANYU14}.  This suggests that if recycling of continental or oceanic crust is the source of the light Si isotope enrichment observed in the HIMU and Iceland NRZ OIB, the material must be geochemically modified either prior to or during subduction.

There are a number of processes acting on oceanic crust that could modify its Si isotope composition, including high-temperature hydrothermal alteration at seafloor spreading centers, low-temperature seafloor alteration, and metasomatism and metamorphic dehydration during subduction \citep{HOFMANN03,STRACKE03,MANNING04}. The preferential loss of heavy Si isotopes to the interacting fluid would result in isotopically light crustal material, which could lead to heterogeneities in the mantle as observed in OIB. Although a high-precision investigation of the Si isotope composition of altered and subducted lithologies (e.g., ophiolite sequences) does not currently exist in the literature, there is some evidence for preservation of initial Si isotope compositions during high-temperature metamorphism and metasomatism \citep{ANDRE06,SAVAGE13c}. Additionally, island arc basalts (IAB) display Si isotope compositions similar to MORB \citep{SAVAGE10}, so processing during subduction may preserve primary Si isotope compositions to the extent that the signature of interacting fluids is not communicated to the island arc. Altered oceanic crust displays enrichments in light Si isotopes relative to MORB \citep{AN16}. It is therefore possible that oceanic crust has experienced Si isotope fractionations due to seafloor and hydrothermal alteration, and this composition is preserved during subduction. However, a systematic study of the behavior of Si isotopes during metamorphic dehydration is necessary to ascertain the effects of subduction on mantle inputs.

In contrast to oceanic and continental crust, oceanic sediments have widely variable compositions and can be significantly enriched in light Si isotopes compared to typical igneous rocks. This suggests that the incorporation of oceanic sediments with isotopically light Si in the OIB plume sources is one viable mechanism for creating the light Si isotope enrichments observed in OIB. Previous measurements of Si isotopes in silicified and precipitated materials such as cherts show a large range in Si isotope composition, with $\updelta$\textsuperscript{30}Si values down to $-$4\textperthousand{} or more \citep{ROBERT06,BOORN07,BOORN10,CHAKRABARTI12}, but sedimentary material with extremely fractionated Si isotope compositions (potentially including clays and cherts) would be diluted by other subducting material. Additionally, much of the sedimentary material that enters subduction zones may be removed before entering the convecting mantle by forearc scraping or through arc volcanism \citep{CLIFT04}; in general, OIB sources are believed to incorporate at most a few percent of sediments \citep{CHAUVEL07,JACKSON07a}. 

For these reasons it is difficult to quantify the Si isotope composition of subducting material that enters the mantle. The potential contribution of subducted material on the Si isotope signature of OIB can be calculated from a simple mass balance using estimates for the Si isotope composition and fraction of recycled material in the OIB source. Here we consider a subducted package with a bulk $\updelta$\textsuperscript{30}Si of $-$1.0\textperthousand{}, $-$0.75\textperthousand{}, and $-$0.5\textperthousand{}. These values are consistent with plausible compositions of altered oceanic crust \citep{AN16} with or without a small portion of sediment \citep[we consider a few percent of sediments with $\updelta$\textsuperscript{30}Si values down to $-$4\textperthousand{} corresponding to compositions of silicified and precipitated materials;][]{ROBERT06,BOORN07,BOORN10,CHAKRABARTI12}. Results of the mass balance calculations are shown in Figure 5. These calculations suggest that the relatively light Si isotope enrichment observed in the HIMU and Iceland NRZ basalts can be explained by the incorporation 25\% of recycled altered oceanic crust or 10-25\% altered oceanic crust plus a few percent of sediments in OIB magmas. Using a similar mass balance shows that 1-2\% sediment added to peridotite mantle is unlikely to measurably shift the $\updelta$\textsuperscript{30}Si of the mixture, which can explain the MORB-like $\updelta$\textsuperscript{30}Si values of EM-1 and EM-2 type OIB despite their hosting some recycled marine sediment. The addition of more than a few percent of sediments in the sources of OIB, particularly HIMU-type, is inconsistent with Pb and Sr isotopic constraints and trace-element ratios unless sediments are significantly modified during subduction zone processing \citep{CHAUVEL07,ROY95}. Sediments constitute a thin veneer that overlies oceanic crust, so that the subduction package is dominated by basaltic oceanic crust and lithospheric mantle.  For this reason, it is more likely that the recycled material that can reproduce the $\updelta$\textsuperscript{30}Si signatures of HIMU and Iceland NRZ OIB is primarily altered `basaltic' oceanic crust with a bulk composition of $\updelta$\textsuperscript{30}Si \textless\  $-$0.5\textperthousand{} that contributes melt to generate $\sim$25\% of the mass of the lavas. Furthermore, it should be noted that the mass fraction of recycled material in the mantle source is unlikely to be the same as the mass fraction present in the plume source melts because recycled lithosphere in the form of eclogite preferentially melts compared to ambient peridotitic mantle, potentially leading to pyroxenite-peridotite mixtures in OIB mantle sources \citep[e.g.,][]{YAXLEY98,SOBOLEV07,DAY09,HERZBERG14}.

\begin{figure}
\centering
\includegraphics[totalheight=3in]{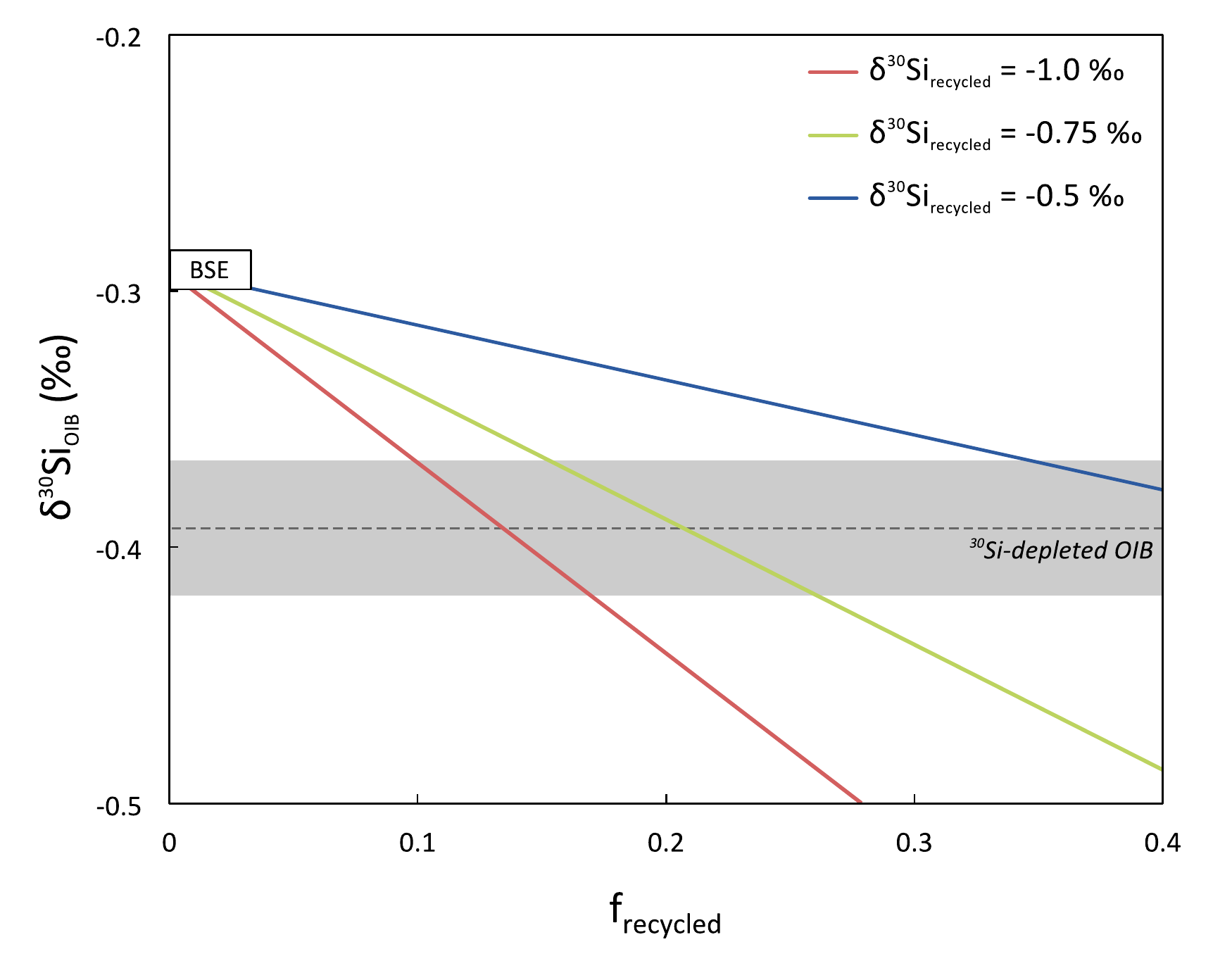}
\caption[Plot of OIB $\updelta$\textsuperscript{30}Si as a function of the fraction of recycled component.]{Plot of OIB $\updelta$\textsuperscript{30}Si as a function of the fraction of recycled component present in the melt derived from the mantle source. The dashed line and shaded box represent the average Si isotope composition ($\pm$ 2 se) of the light-isotope enriched OIB measured in this study (Iceland, Mangaia, Cape Verde). The BSE reservoir is considered to have a Si concentration [Si] = 0.21 wt.\% and $\updelta$\textsuperscript{30}Si = $-$0.29 \citep{SAVAGE14}. The recycled component is modeled with a nearly identical bulk [Si] = 0.23 wt.\% but variable $\updelta$\textsuperscript{30}Si values. Here $-$0.50, $-$0.75, and $-$1.00 are used as illustrative $\updelta$\textsuperscript{30}Si values for the bulk recycled component; these values are consistent with altered oceanic crust \citep{AN16} or a mixture of altered oceanic crust with a few percent of sediments ($\updelta$\textsuperscript{30}Si values between $-$2\textperthousand{} and $-$4\textperthousand{}; see text for details). The Si isotope compositions of the anomalous OIB localities are consistent with the incorporation of 10-30\% recycled material in the plume source melt.} 
\end{figure}

A key indication that the Si isotope variability that potentially arises from recycling of subducted oceanic crust and lithospheric mantle lies in the difference in Si isotope composition between lavas from the Canary Islands of La Palma and El Hierro. These two islands have previously been found to have contrasting O isotope compositions, with relatively lower $\updelta$\textsuperscript{18}O values in La Palma samples compared to El Hierro samples \citep{DAY09}. Furthermore, the coupled O and Os isotope systematics of La Palma and El Hierro reflect different proportions of recycled oceanic crust and lithospheric mantle in the plume sources \citep{DAY09,DAY10}.  Specifically, the relatively lower $\updelta$\textsuperscript{18}O values in La Palma are generated by a contribution of recycled oceanic gabbroic crust \citep{DAY09}. This suggests that a single process may be responsible for generating the low $\updelta$\textsuperscript{18}O and low $\updelta$\textsuperscript{30}Si values observed in oceanic basalts, although further work is necessary to confirm this relationship. These observations indicate that the Si isotope system may be an important tool for investigating the processes responsible for generating stable isotopic variability present in OIB. 


\section{Conclusions}

To a first order, OIB exhibit homogeneous Si isotope compositions generally in agreement with estimates for the $\updelta$\textsuperscript{30}Si value of BSE \citep[$-$0.29 $\pm$ 0.07\textperthousand{}, 2 sd;][]{SAVAGE14}. However, some systematic variations are present; some HIMU-type and NRZ  basalts exhibit light Si isotope enrichments relative to other OIB, MORB, and the estimate for the BSE. In contrast, EM-type OIB do not show any systematic Si isotope variations. One possible cause of the light Si isotope enrichments in OIB is primitive Si isotope heterogeneity created by bridgmanite crystallization at the base of the magma ocean during the silicate differentiation of the Earth as suggested by \citet{HUANG14}, although the potential to preserve such a reservoir over the history of the Earth is debated. An alternative explanation is that subduction of material that is incorporated into hotspot sources may give rise to observed heterogeneities. The Si isotope variability in the HIMU and Iceland NRZ can be explained by the presence of $\sim$25\% recycled altered oceanic crust and lithospheric mantle in the plume source. 
	
A reservoir of isotopically light Si in the Earth's mantle would have implications for the calculated amount of Si in the core. Silicon has long been proposed as a potential light element in the core to account for the density deficit relative to pure FeNi, and an estimate of the Si content of the Earth's core was one of the early applications of the Si isotope system to high temperature geochemistry \citep{GEORG07,FITOUSSI09}. Debate still persists about the degree of offset between the Si isotope composition of the Earth's mantle and that of chondrites (assumed to represent the bulk Earth composition), resulting in widely varying estimates of the Si content of the Earth’s core from $\sim$1 wt.\% to more than 10 wt.\%, although high amounts of Si in the terrestrial core are likely irreconcilable with other geochemical and geophysical constraints \citep[e.g.,][]{MCDONOUGH03,BADRO07,GEORG15}. However, a reservoir of isotopically light Si in the mantle may be indicated by these new measurements of OIB. This would lower the amount of Si in the core required by mass balance, driving the calculated Si content of the core based on Si isotopes closer to that estimated by geophysical means.

\section*{Acknowledgments}

We thank the three anonymous reviewers as well as the Associate Editor Fangzhen Teng for their thorough reviews and comments, which have greatly improved this paper. EP thanks the Chateaubriand STEM fellowship program for funding. FM thanks the European Research Council under the European Community's H2020 framework program/ERC grant agreement \#637503 (Pristine) and the Agence Nationale de la Recherche for a chaire d'Excellence Sorbonne Paris Cit\'{e}́ (IDEX13C445) and for the UnivEarthS Labex program (ANR-10-LABX-0023 and ANR-11-IDEX-0005-02). PS thanks the support of the Marie Curie FP7-IOF fellowship ``Isovolc''.

\cleardoublepage

\end{document}